\documentclass[prd,preprint,aps,amssymb,tightenlines]{revtex4}

\usepackage{amsmath}

\begin{document}

\title{Possible constraints on string theory in closed space\\
with symmetries}

\author{Atsushi Higuchi}

\affiliation{Department of Mathematics, University of York,
Heslington, York YO10 5DD, UK\\ email: ah28@york.ac.uk}

\date{November 23, 2005}

\

\begin{abstract}
It is well known that certain quadratic constraints have to be imposed on
linearized gravity in closed space with symmetries.  We review this
phenomenon and discuss one of the constraints which arise in linearized
gravity on static flat torus in detail.  Then we point out that
the mode with negative kinetic energy, 
which is necessary for satisfying this constraint,
appears to be missing in the free bosonic string
spectrum.
\end{abstract}

\maketitle

\section{Introduction}
(Super)string theory is the leading candidate for a unified theory including
gravity.  In particular, 
it contains and generalizes Einstein's general 
relativity~\cite{hig-yoneya1,hig-yoneya2,hig-scherk}.
Therefore, it is natural to expect that the theory incorporates diffeomorphism
invariance.  However, this invariance is not manifest in the
perturbative definition of string theory starting from 
non-interacting string. 
Now, it is well known that a solution of linearized Einstein equations (with
or without matter fields) in compact background space with Killing symmetries
cannot be extended to an exact solution unless the linearized solution 
satisfies certain quadratic constraints~\cite{hig-brill,hig-fischer}. 
This phenomenon, called linearization
instability, is a consequence of diffeomorphism invariance of the
full theory.
(This fact can
be seen most clearly in the quantum context.)  
Therefore, one may gain some insight into how diffeomorphism
invariance is incorporated in string theory by investigating the way
linearization instabilities manifest themselves.

In this article we review the phenomenon of linearization instability
in general relativity
with emphasis on the case with static flat torus space.  In particular,
we point out that in this space
a mode with negative kinetic term is essential in satisfying one of the
constraints and that this mode seems to be missing in the spectrum of free
bosonic string theory.  The rest of the article is organized as follows. 
In Section 2 the phenomenon of linearization instability in 
classical and quantum general relativity is reviewed.  In Section 3 
one of the constraints occurring in flat torus space is discussed in
detail and the importance of a mode with negative kinetic term is emphasized.
In Section 4 it is pointed out that this mode is absent in a seemingly natural
treatment of the zero-momentum sector of closed bosonic string in this
space.  In Section 5 a summary of this article is given. The metric signature
is $(-++\cdots +)$ throughout this article.
 
\section{Linearization instabilities in general relativity}

Consider classical general relativity with any bosonic 
matter fields.  Suppose we want to find a solution in this theory order 
by order in perturbation theory starting from a (globally-hyperbolic)
background spacetime
satisfying the vacuum Einstein equations $R_{ab} = 0$.  To do do so we write
the metric $g_{ab}$ and the matter fields $\phi_i$ as
\begin{eqnarray}
g_{ab} & = & g^{(0)}_{ab} + h^{(1)}_{ab} + h^{(2)}_{ab} + \cdots\,, 
\nonumber \\
\phi_i & = & \phi_i^{(1)} + \phi_i^{(2)}+ \cdots\,, \nonumber
\end{eqnarray}
where $g^{(0)}_{ab}$ is the background metric and where $h^{(k)}_{ab}$
and $\phi_i^{(k)}$ are the fields obtained as the $k$-th order approximation.
(The fields $\phi_i$ are assumed to vanish at zero-th order for simplicity.)
The first-order 
approximation $(h^{(1)}_{ab},\phi_i^{(1)})$ corresponds to non-interacting
waves in the background spacetime.
The second-order perturbation of the metric,
$h^{(2)}_{ab}$, can be regarded as the gravitational field generated by
the free fields $h^{(1)}_{ab}$ and $\phi_i^{(1)}$.  

Let the stress-energy
tensor of the fields $h^{(1)}_{ab}$ and $\phi_i^{(1)}$ in the background
spacetime with metric $g^{(0)}_{ab}$ be $T^{(1)}_{ab}$.  We note first
that the linear contribution to the Einstein tensor 
$$
E_{ab} = R_{ab} - \frac{1}{2}g_{ab}R
$$ 
with $g_{ab} = g_{ab}^{(0)} + h_{ab}$ is 
\begin{eqnarray}
E_{ab}^{(L)}(h) & = & \frac{1}{2} ( \nabla_c \nabla_b {h^{c}}_a
+ \nabla_c \nabla_a {h^{c}}_{b} - \nabla_c \nabla^c h_{ab}
- \nabla_a \nabla_b {h^{c}}_c) \nonumber \\
 & & - \frac{1}{2}g_{ab}^{(0)}( \nabla_c \nabla_d h^{cd} - \nabla_c\nabla^c
{h^{d}}_d)\,. \nonumber
\end{eqnarray}
Here the covariant derivatives are compatible with the metric $g^{(0)}_{ab}$
and indices are raised and lowered by this metric. 
The field $h^{(2)}_{ab}$ must satisfy
\begin{equation}
E_{ab}^{(L)}(h^{(2)}) = \kappa T^{(1)}_{ab}\,,  \label{hig-linear}
\end{equation}
where $\kappa$ is a constant.
The stress-energy tensor $T^{(1)}_{ab}$ is divergence-free, i.e.,
$\nabla^a T^{(1)}_{ab} = 0$, if the linear equations of motion are satisfied.
On the other hand the equation
\begin{equation}
\nabla^a E_{ab}^{(L)}(h) = 0 \label{hig-bianchi}
\end{equation}
holds for {\em any} $h_{ab}$.  This is a consequence of the Bianchi
identity $\tilde{\nabla}^{a} E_{ab} = 0$, 
where $\tilde{\nabla}_a$ is the covariant derivative
compatible with the full metric $g_{ab}$.
For this reason Eq.~(\ref{hig-bianchi})
is called the background Bianchi identity.

Now, suppose that there is a Killing vector field $X^a$ satisfying
$$
\nabla_a X_b + \nabla_b X_a = 0\,.
$$
Then, it is easy to verify that the current $j_X^{a} \equiv T^{(1)ab}X_b$ 
is conserved.  The corresponding conserved Noether charge is given by
$$
Q_X \equiv \int_{\Sigma} d\Sigma\, n_a j_X^{a}\,,
$$
where the integration is over any Cauchy surface $\Sigma$
and $n_a$ is the unit normal
to the Cauchy surface. (Since $Q_X$ comes from a stress-energy tensor
of the {\em free} fields $h_{ab}^{(1)}$ and $\phi_i^{(1)}$, it is quadratic
in these fields.)
If the vector $X^a$ is a 
time-translation Killing vector,
then the charge $Q_X$ is nothing but the energy.  If it is a space-translation
Killing vector, then $Q_X$ is a component of the momentum.  We note that 
$$
E^{(L)ab}(h)X_b = \frac{1}{2}\nabla_b K^{ab}(h)\,,
$$
where $K_{ab}(h)$ is an anti-symmetric tensor given by
\begin{eqnarray}
K_{ab}(h) & = & X_a \nabla_b {h^c}_c - X_b \nabla_a {h^c}_c 
+ X^c \nabla_a h_{bc}
- X^c \nabla_b h_{ac} \nonumber \\
& & + X^c \nabla_a h_{bc} - X^c \nabla_b h_{ac}
+ h_{ca}\nabla_b X^c - h_{cb} \nabla_a X^c\,.  \nonumber
\end{eqnarray}
Hence, the integral of $E^{(L)ab}(h)X_b$ over the Cauchy surface
can be expressed as a surface integral as
$$
\int_{\Sigma} d\Sigma\,n_a E^{(L)ab}(h)X_b
= \frac{1}{2}\int_{\partial\Sigma} dS\, n_a r_b K^{ab}(h)\,, 
$$
where $\partial\Sigma$ is the ``boundary" of the Cauchy surface at infinity and
$r_a$ is the unit vector normal to the boundary along the Cauchy surface.
By using this expression and Eq.~(\ref{hig-linear}) one can write
the Noether charge $Q_X$ as a surface integral:
\begin{equation}
Q_X = \frac{1}{2\kappa} 
\int_{\partial\Sigma} dS\, n_a r_b K^{ab}(h^{(2)})\,. \label{hig-surface}
\end{equation}
In asymptotically-flat spacetime this equation allows us to express energy
and momentum of an isolated system as surface integrals at spacelike 
infinity~\cite{hig-ADM}.

Now, suppose that the Cauchy surface is compact, i.e., that the space is
``closed".  Then, the right-hand side
of Eq.~(\ref{hig-surface}) must vanish for any $h_{ab}$ because there is
no surface term. 
Hence,
\begin{equation}
Q_X  = 0\,.  \label{hig-QX}
\end{equation}
Thus, the conserved charge $Q_X$ is constrained to vanish.  Note that this
constraint cannot be derived from the 
linearized theory alone.  It
arises in the full theory when we try to find the correction to the linear 
theory.  Solutions of the linearized field equations are not extendible to
exact solutions unless they satisfy this constraint. (The
background spacetime here is said to be linearization unstable because of the
existence of spurious solutions to the linearized equations.  The constraint
(\ref{hig-QX}) is sometimes called a linearization stability condition.) 

Although we will concentrate on classical theory, it 
is interesting to note what the constraint (\ref{hig-QX}) implies in
quantum theory.  In the Dirac quantization, constraints are imposed on the 
physical states.  Thus, the quantum version of (\ref{hig-QX}) reads
\begin{equation}
{\cal Q}_X|{\rm phys}\rangle = 0\,,  \label{hig-qQX}
\end{equation}
where $|{\rm phys}\rangle$ is any physical state and ${\cal Q}_X$ 
is the quantum operator corresponding to the conserved Noether charge $Q_X$.
Since the operator ${\cal Q}_X$ generates the spacetime symmetry associated
with the Killing vector field $X^a$, the constraint (\ref{hig-qQX}) implies
that all physical states must be invariant under this spacetime 
symmetry~\cite{hig-moncrief}.  This requirement might seem absurdly strong
at first sight. For example, in linearized gravity in de~Sitter spacetime 
{\it all} physical states are required to be de~Sitter
invariant.\footnote{The vacuum state is the only
de~Sitter invariant state if one insists on using the original Fock space 
of linearized gravity, but
one can construct infinitely many invariant states by using a different
Hilbert space~\cite{hig-higuchi}.} 
However, in the (formal) Dirac quantization of full general relativity, the
states are (roughly speaking)
required to be diffeomorphism invariant.  The constraint (\ref{hig-qQX})
can be interpreted to be inforcing
the part of the diffeomorphism invariance of the physical
states that has not been broken by the background metric.

\section{The Hamiltonian constraint 
of linearized gravity on flat torus}

In this section we discuss linearized gravity in
static flat $(D-1)$-dimensional torus space with
all directions compactified.  
This spacetime has space- and
time-translation invariance.  Therefore, the energy and momentum of linearized
gravity are 
conserved and are both constrained to vanish.
Below we concentrate on 
the linearization stability condition which requires that
energy be zero since it will be important later in the discussion of
string theory.
We find that there is a mode with negative kinetic term
and that there would be no excitation as a result of the linearization
stability condition if it were not for this mode. 
We consider only pure gravity for simplicity. 

Let us impose the standard (``Lorenz" or Hilbert) gauge condition 
\begin{equation}
\partial_a h^{ab} = \frac{1}{2}\partial^b h\,, \label{hig-DeDonder}
\end{equation}
where $h = {h^c}_c$. 
Then the Hamiltonian density reads
$$
{\cal H}  =  \frac{1}{4}\left[ \partial_t \tilde{h}_{ab}\partial_t
\tilde{h}^{ab} + \partial_i \tilde{h}_{ab}\partial^i \tilde{h}^{ab}\right]
 - \frac{D-2}{4D}\left[ (\partial_t h)^2 + \partial_i h \partial^i h\right]
\,,
$$
where $\tilde{h}_{ab} = h_{ab} - \frac{1}{D}g_{ab} h$ is the traceless part
of $h_{ab}$.  The index $i$ runs from $1$ to $D-1$, i.e., it is a spacelike
index.  The field equations are simply
$$
\Box \tilde{h}_{ab} = 0\,, \ \ \  \Box h = 0\,.
$$
The modes with nonzero momentum ${\bf k}$ are proportional to 
$e^{-ik^0 t + i{\bf k}\cdot{\bf x}}$,
where $(k^0)^2 - {\bf k}^2 = 0$. On the other hand, the
modes with ${\bf k} = 0$ take the form
$$
\tilde{h}_{ab}\,,\ h\, \propto A t + B\,,
$$
where $A$ and $B$ are constants.\footnote{Note that the energy 
corresponding to these
modes would be infinite for $A \neq 0$  if the space were not compactified.
This is why these modes would not be present in uncompactified space.}

The Hamiltonian can be written as
$$
H = \int d^{D-1}{\bf x}\, {\cal H} = H_0 + H'\,,
$$
where $H_0$ is the energy in the modes with ${\bf k} = 0$ and where $H'$
is the energy in the modes with ${\bf k} \neq 0$.  For the modes with
${\bf k} \neq 0$ the trace $h$ can be gauged away and the physical modes 
have the form
$$
\tilde{h}_{ab} \propto H_{ab} e^{-ik^0 t + i {\bf k}\cdot{\bf x}}\,,
$$
where $H_{ab}$ is a constant symmetric tensor satisfying $H_{tb} = 0$,
${H^i}_i = 0$ and $k^i H_{ij} = 0$.  Then we can easily see that
$H' \geq 0$.  The situation is rather different for the modes
with ${\bf k} = 0$. 
Since these
modes are constant in space, they satisfy 
$\partial_i \tilde{h}_{ab} = \partial_i h = 0$.  Hence, the 
conditions coming from (\ref{hig-DeDonder}) are 
$\partial_t \tilde{h}_{ti} = 0$ and
$$
\partial_t \tilde{h}_{tt} = - \frac{D-2}{2D}\partial_t h\,.
$$
Let us write 
\begin{eqnarray}
\tilde{h}_{ab} & = & \tilde{h}_{ab}^{(0)} + \tilde{h}_{ab}'\,, \nonumber \\
h & = & h^{(0)} + h'\,,  \nonumber
\end{eqnarray}
where $\tilde{h}_{ab}^{(0)}$ and $h^{(0)}$ are the zero-momentum parts
of $\tilde{h}_{ab}$ and $h$.  Then the zero-momentum Hamiltonian $H_0$ is
given by
$$
H_0 = \int d^{D-1}{\bf x}\,\left[
\frac{1}{4}\partial_t \tilde{h}_{ij}^{(0)}\partial_t \tilde{h}^{(0)ij}
- \frac{D^2-4}{8D} (\partial_t h^{(0)})^2\right]\,.
$$
Notice that the trace mode $h^{(0)}$ has a negative kinetic term.

Since the Hamiltonian is the Noether charge corresponding to the 
time-translation symmetry of the background spacetime, the 
discussion in the previous section shows that 
$$
H=H_0 + H' = 0\,.
$$
The solutions
of the linearized equations which do not satisfy this condition cannot be
extended to exact solutions. This equation can be re-expressed as
\begin{equation}
- \frac{D^2 - 4}{8D} \int d^{D-1}{\bf x}\, (\partial_t h^{(0)})^2
+ H'' = 0\,,  \label{hig-cosmology}
\end{equation}
where
$$
H'' = H' + \frac{1}{4}\int d^{D-1}{\bf x}\,\partial_t \tilde{h}_{ij}^{(0)}
\partial_t \tilde{h}^{(0)ij} \geq 0\,.
$$
Now, the quantity $\frac{1}{2}h^{(0)}V$, where 
$V$ is the volume of the background space,
is the change in the volume of the space.  Hence,
Eq.~(\ref{hig-cosmology}) relates the expansion/contraction rate of
space to the energy due to the excitation of the system.  
In fact this equation is 
the linearized version of a familiar equation in cosmology.  Notice that
the trace mode $h^{(0)}$ plays a vital role in satisfying 
Eq.~(\ref{hig-cosmology}).  If this mode were absent, 
Eq.~(\ref{hig-cosmology}) would imply that there were no
excitations on flat torus compactified in all directions.

\section{Massless sector of bosonic string in the position representation}

Massless excitations of closed string include 
gravitons,
i.e., linearized gravity is present among the modes of free
closed bosonic string in Minkowski spacetime.\footnote{This fact goes beyond
the linearized level as is well 
known\cite{hig-yoneya1,hig-yoneya2,hig-scherk}.}
This fact is one of the most important features of string theory as a unified
theory.  It is natural to expect that this feature persists in string theory
in static
flat torus compactified in all directions.  Therefore, the total energy and
momentum in string (field) theory are expected to vanish in this spacetime. 
We also expect that there is a mode with negative kinetic term
among the closed-string modes so that the linearization stability condition
(\ref{hig-cosmology}) can
be satisfied by non-vacuum states (in string field theory). 
However, we will find in the
``old covariant approach" that there is
no massless string excitation which corresponds to the 
zero-momentum mode $h^{(0)}$ with negative kinetic energy
if we treat the zero-momentum modes in a way which seems
most natural.

Let us start with a discussion of open string in flat $(D-1)$-dimensional
torus.  The massless 
states in the old covariant approach are denoted by 
$$
\alpha_{-1}^{a}|0;p\rangle\,,
$$
where the state $|0;p\rangle$ with momentum $p^a$ has no string excitation
(see, e.g., Ref.~\cite{hig-witten}). 
The creation
operator $\alpha_{-1}^a$ creates the lowest harmonic-oscillator mode on
the string 
in the $a$-direction and the annihilation operator $\alpha_1^a$ annihilates
it.  As is well known, the physical state conditions lead to $p^2 = 0$ and
$p\cdot \alpha_1|{\rm phys}\rangle = 0$, where
$[\alpha_{1}^a, \alpha_{-1}^b] = g^{ab}$ and 
$p\cdot \alpha_1 \equiv p_a \alpha_{1}^a$.  [Here, 
$g_{ab}={\rm diag}(-1,1,1,\ldots, 1)$.] 
Let us consider a wave-packet state
$$
|\psi\rangle =\int\frac{d^D p}{(2\pi)^D}\,
\hat{A}_a(p)\alpha_{-1}^a|0;p\rangle\,,
$$
where $\hat{A}_a(p)$ is a function of $p^a$.
The physical state conditions then read 
$p^2\hat{A}_a(p) = 0$ and
$p^a \hat{A}_a(p) = 0$.
Now, define the (spacetime) position representation of this wave packet as
$$
A_a(x) = \int \frac{d^Dp}{(2\pi)^D}\, \hat{A}_a(p)e^{-ip\cdot x}\,.
$$
Then the physical state conditions become
$\Box A_a = 0$ and
$\partial^a A_a  = 0$.
Thus, we recover the equations satisfied by a non-interacting $U(1)$
gauge field in the Lorenz gauge.  
The zero-momentum modes in flat $(D-1)$-dimensional torus
satisfy
$$
\partial_t A_{t} = 0\,,\ \ \partial_t^2 A_{i} = 0\,.
$$
These imply that $A_t = {\rm const}$ and $A_i = E_i t + A_i^{(0)}$.  The
constant $A_t$ can be gauged away, but the constants $E_i$ (the electric field)
and $A_i^{(0)}$ represent physical degrees of freedom. 

Next, 
we will apply the above procedure to a closed string on static flat torus
and examine whether or not there is 
a mode with negative kinetic term.
The massless excitations of a closed bosonic string are
$$
\alpha_{-1}^a \tilde{\alpha}_{-1}^b|0;p\rangle\,.
$$
The operator $\alpha_{-1}^a$ ($\tilde{\alpha}_{-1}^a$) creates the lowest 
left-moving (right-moving) mode on the string in the $a$-direction,
and the operator $\alpha_1^a$ and $\tilde{\alpha}_1^a$ annihilate them.
The physical state conditions lead to 
$p^2 = 0$ and
$p\cdot \alpha_1|{\rm phys}\rangle 
= p\cdot\tilde{\alpha}_1|{\rm phys}\rangle = 0$,
where $[\alpha_1^a,\alpha_{-1}^b] 
= [\tilde{\alpha}_1^a,\tilde{\alpha}_{-1}^b] = g^{ab}$.  We again
consider a wave-packet state 
$$
|\Psi\rangle = \int \frac{d^D p}{(2\pi)^D}\, \hat{H}_{ab}(p)
\alpha_{-1}^a\tilde{\alpha}_{-1}^b|0;p\rangle\,.
$$
(Note here that the tensor $\hat{H}_{ab}(p)$ is not necessarily symmetric.)
The physical state conditions read $p^2 \hat{H}_{ab}(p) = 0$ and
$p^a \hat{H}_{ab} = p^b \hat{H}_{ab} = 0$.  In the spacetime position
representation, 
$$
H_{ab}(x) = \int \frac{d^D p}{(2\pi)^D}\,
\hat{H}_{ab}(p)e^{-ip\cdot x}\,,
$$
the physical state conditions are 
$\Box H_{ab} = 0$ and
\begin{equation}
\partial^a H_{ab} = \partial^b H_{ab} = 0\,. \label{hig-Hconstraint}
\end{equation}

The equation
$\Box H_{ab}= 0$ naturally comes from the following Lagrangian
density:
\begin{equation}
{\cal L} = - \frac{1}{4}\partial_a H_{bc} \partial^a H^{bc}\,.
\label{hig-naive}
\end{equation}
The constraints (\ref{hig-Hconstraint}) can be imposed by hand.
One finds the modes corresponding to gravitons, anti-symmetric tensor
particles and dilatons in the nonzero momentum sector of this theory as
in Minkowski spacetime.  
The constraints (\ref{hig-Hconstraint}) for the zero-momentum sector read
$$
\partial_t H_{ta} = \partial_t H_{at} = 0
$$
for all $a$.  The energy in the zero-momentum sector is
$$
E_0 = \frac{1}{4} \int d^{D-1}{\bf x}\, \partial_t H_{ij}\partial_t H^{ij}\,,
$$
where $i, j = 1,2,\cdots D-1$.  There is no mode with negative kinetic term
in this expression, and $E_0$ is positive definite.  Thus, the negative-energy 
mode, which is necessary for non-vacuum states to satisfy the constraint
(\ref{hig-cosmology}), does not appear in a seemingly natural position
representation of the massless sector of closed bosonic string.

\section{Summary}

In this article, we reviewed the fact that quadratic constraints arise in
linearized gravity if the background spacetime allows Killing symmetries and
has compact Cauchy surfaces.  This implies that the total energy and 
momentum in free string (field) theory should be constrained to vanish in
flat torus space with all directions compactified.
We examined one of these constraints
in linearized gravity
in this space, emphasizing that a
mode with negative kinetic energy is essential 
in satisfying this constraint.  Then we analyzed free
closed bosonic string theory
in this space and found that this mode does not appear
in a seemingly natural treatment of the massless sector.

It is possible that the Lagrangian density (\ref{hig-naive}) is wrong, and 
a more careful analysis may 
lead to a Lagrangian density describing the usual
linearized gravity, anti-symmetric tensor gauge field and dilaton
scalar field after all.  It will be interesting to see how this can be
achieved.  The situation is rather puzzling, 
however, because string theory
is formulated in terms of a physical object, i.e. a string, and 
does not seem to allow any mode with negative kinetic energy.


\end{document}